\documentclass[12pt]{iopart}
\usepackage{graphicx}
\usepackage[colorlinks=true, linkcolor=blue, citecolor=blue, urlcolor=blue]{hyperref}
\usepackage{iopams} 
\begin{document}

\title[Coherent microwave-to-optical transduction with Yb:YSO]{Coherent microwave-to-optical transduction with Yb:YSO spins strongly coupled to a 3D resonator}

\author{Ujjwal Gautam, Nasser Gohari Kamel, Sourabh Kumar and Daniel Oblak}

\address{\label{InstituteQuantumCalgary}Institute for Quantum Science and Technology, University of Calgary, Calgary, AB, Canada T2N 1N4 \\
\label{PhysicsCalgary}Department of Physics and Astronomy, University of Calgary, Calgary, AB, Canada T2N 1N4}
\ead{doblak@ucalgary.ca}

\vspace{10pt}
\begin{indented}
\item[]\today
\end{indented}

\begin{abstract}
Microwave-to-optical quantum transducers are essential for entangling remote superconducting qubits. Among the available transduction platforms, ensembles of Er$^{3+}$ and Yb$^{3+}$ ions doped into solids have emerged as leading candidates. While external magnetic fields are needed to split the Zeeman levels of erbium ions and enable a microwave--qubit interface, superconducting qubits suffer decoherence in such fields. In contrast, ytterbium ions exhibit zero-first-order Zeeman transitions and large hyperfine splittings at zero magnetic field (when doped into inorganic crystals). Owing to its long optical and spin coherence times, Yb:YSO has been widely used as a quantum memory, yet its potential for quantum transduction remains largely unexplored. Investigating this material could enable the integration of quantum memory and transduction in a single platform.
Here, we demonstrate microwave-to-optical transduction in the continuous-wave regime using a 5\,ppm doped Yb:YSO crystal. The internal transduction efficiency is $2\times10^{-8}$ with a bandwidth of 200\,kHz, achieved using a 3D loop-gap microwave resonator and a single-pass optical configuration. We explore all the ground states that form a V-type three-level system with the first and second optical excited states and assert the use of the ground state, which provides the highest efficiency and isolated optical transition. We further establish strong spin-microwave coupling from avoided crossing measurements. With a strong microwave drive to saturate the spin transition, we estimate the spin population pumped into the excited state, which is close to the simulated value. Finally, we calculate target parameter values for maximum efficiency with our system and suggest using 50\,ppm doped Yb:YSO crystal. With the calculated target parameters, the internal transduction efficiency is predicted to reach up to $10^{-4}$ in the current 3D loop-gap resonator.
\end{abstract}

\section{Introduction}\label{sec:Intro}

Superconducting qubits are a leading candidate for quantum computing~\cite{acharya-2024}, but their scalability remains a challenge due to the physical size and cryogenic cooling required to reach the necessary scale~\cite{megrant-2025}. Distributed quantum computing~\cite{cirac-1999} is a proposed solution to increase the number of qubits in a superconducting quantum computer. In this approach, several nodes of superconducting quantum circuits (SQCs) are interconnected via quantum interconnects to multiply the number of qubits available for quantum computation.
A quantum interconnect working at microwave frequencies (1-10\,GHz) and at room temperature will inevitably introduce excessive thermal noise (blackbody radiation) on top of the quantum signal, rendering the channel useless for quantum applications. To overcome this thermal noise, researchers have explored cryogenic transmission lines between two nodes of SQCs~\cite{magnard-2020, Salari2024}. However, such cryogenic transmission lines are difficult to scale-up and integrate with a large number of nodes over long distances.

A quantum microwave-to-optical (M2O) transducer~\cite{xie-2025,han-2021, kumar-2019} offers an alternative path to build the quantum interconnections, in which the microwave quantum signals from the SQCs are faithfully converted into flying optical quantum signals for noise-free and low-loss transmission over room temperatures. The quantum M2O transducers should add minimal noise to the output optical signal while operating with large bandwidth and high efficiencies. Moreover, to avoid decoherence in the SQCs, the locally placed transducers must operate at zero external magnetic field~\cite{kakuyanagi-2007}. 

In essence, M2O transduction is a non-linear process of converting photons with orders of magnitude energy difference~\cite{han-2021}. Several platforms have been explored for transduction e.g. opto-mechanics~\cite{higginbotham-2018,weaver-2023}, electro-optics~\cite{sahu-2022} and rare-earth ion ensembles~\cite{xie-2025,  fernandez-gonzalvo-2019, king-2024, rochman-2023}. These demonstrations have shown promising efficiencies, but feature trade-offs with other figures of merit such as bandwidth and noise. Therefore, exploration of the perfect platform for quantum transduction is still ongoing.

Among rare-earth ensembles, Er\textsuperscript{3+}~\cite{fernandez-gonzalvo-2019} and Yb\textsuperscript{3+}~\cite{xie-2025} ions doped in various host crystals have mostly been explored for transduction. Although, Er\textsuperscript{3+} ions feature an attractive telecom-wavelength optical transition -- useful for integration with the existing optical fiber infrastructure -- it requires magnetic field to split the Zeeman levels to interact with the microwave field. Therefore, Er-based transducers are not an appropriate choice when operating locally in conjunction with SQCs. On the other hand, Yb\textsuperscript{3+} ion-doped crystals possess hyperfine splittings in the\,GHz frequencies at zero magnetic field, making them ideally suitable to operate locally with the SQCs. Also, Yb\textsuperscript{3+} ions have zero-first-order-Zeeman (ZEFOZ) transition at zero magnetic field, which makes the optical and spin coherence less susceptible to micro- or macroscopic environmental magnetic field fluctuations. 
Recent demonstration of transduction with Yb:YVO has set the benchmark for the highest transduction efficiency ($\sim 1\%$) with rare-earth ions~\cite{xie-2025}. Alternatively, Yb:YSO crystal, due to its long optical and spin coherence times, has been extensively explored for quantum memories applications~\cite{kamel-2025, businger-2022, businger-2020}, and proposed for quantum transducers~\cite{nicolas-2023}. The crystal also has strong branching ratios for isolated optical transitions, which makes it easier to work with.

In this work, we demonstrate continuous wave (CW) M2O transduction with a 5\,ppm doped Yb:YSO crystal and establish strong spin-cavity coupling. The strong spin-cavity coupling is achieved using a 3D loop-gap resonator (LGR) interacting with an excited state spin transition. With a single-pass optical configuration, we report an internal transduction efficiency of $2\times 10^{-8}$ with a bandwidth of 200\,kHz while operating at zero applied magnetic field. We explore different ground levels for optical transitions and assert the use of $\vert 4_g\rangle$ level (see Fig.~\ref{fig:setup} (a)) for further demonstrations of transduction. We also show the saturation of spins at high microwave powers and estimate the number of spins in the interaction mode volume based on the saturating power. Further, we discuss the possible improvements in efficiency using Yb:YSO as the ionic ensemble for transduction.

\section{Setup and Protocol}

\subsection{Setup}
\begin{figure}[t]
\centering
\includegraphics[width=\linewidth]{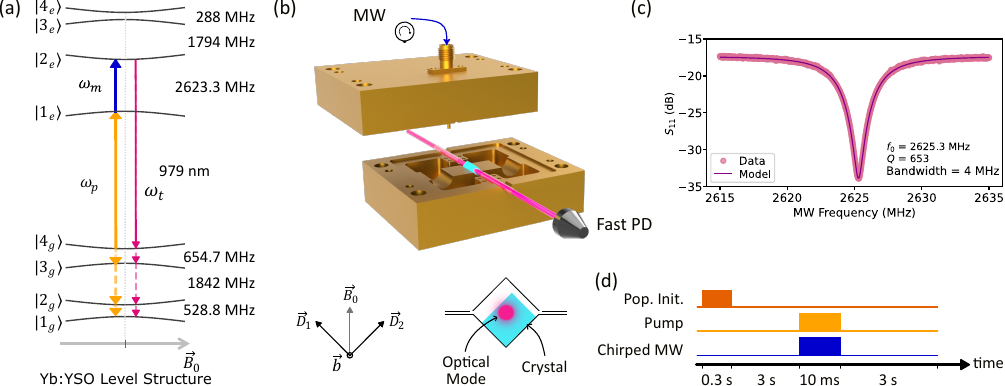}
\caption{(a) Energy level structure of Yb:YSO crystal with the relevant optical and spin transitions used in the experiments. (b) Open view of a 3-loop-2-gap resonator (LGR) with the crystal placed at the center. The microwave is coupled through a single-port SMA antenna pin. The optical field is coupled via GRIN lens and capillary holding ferrules on each side. The output optical field is detected on a fast photodetector. Orientation of the crystal with respect to the external DC magnetic field. Overlap of the optical mode with the crystal showing the interaction region. (c) LGR reflection spectrum fitted to our model showing the relevant fitting parameters. (d) Pulse sequence consisting of population initialization, wait times for population relaxation and heat dissipation, and continuous transduction with pump and chirped microwave pulses.}
\label{fig:setup}
\end{figure}

M2O transduction with atomic system involves three-levels with one spin and two optical transitions. Here, we use the excited state transition, $\vert 1_e\rangle \leftrightarrow \vert 2_e\rangle$ of Yb:YSO for the spin transition at frequency $\omega_m=2623.3$\,MHz as shown with the blue arrow in Fig.~\ref{fig:setup}(a). The transitions connecting these two excited states to any of the ground states provide the two optical transitions, shown as yellow and red arrows in Fig.~\ref{fig:setup}(a). We choose $\vert i_g\rangle \rightarrow \vert 1_e\rangle$ at frequency $\omega_p$ to be the pump transition and $\vert 2_e\rangle \rightarrow \vert i_g\rangle$ to be the transduced optical transition at frequency $\omega_t$, where $i\in \{1,2,3,4\}$. Here, the energy conservation is maintained, $\omega_t = \omega_p + \omega_m$. These optical and spin transitions are ZEFOZ at zero magnetic field. However, we do not apply any external magnetic field to cancel the residual environmental background which we assess to be on the order of a few $\mu$T~\cite{nicolas-2023}.

Our setup consists of a 3D LGR for enhanced microwave interaction with the spin transition and without an optical cavity (see Fig.~\ref{fig:setup}(b)). The design of the LGR is inspired by ref.~\cite{ball2018loop}. It features two cylindrical loops, two fine gaps, and a diamond-shaped cut-out/loop at its center. We keep a 5\,ppm doped Yb:YSO crystal with dimensions $2.5\times2.6\times10$~mm along the $\vec{D_1} \times \vec{D_2} \times \vec{b}$ crystal axis in the diamond loop where it is held and thermalized using vacuum grease. The diamond shape of the central loop is chosen to: (i) thermalize the crystal from two surfaces instead of one in a rectangular design~\cite{ball2018loop}, (ii) naturally orient the crystal perpendicular to the optical and microwave AC magnetic fields in the region, and (iii) provide a homogeneous AC magnetic field region in the optical mode volume highlighted in Fig.~\ref{fig:setup}(b) (bottom). The setup is anchored to the base plate of a dilution fridge at 30\,mK temperature.

The LGR is simulated in COMSOL Multiphysics with the following constraints: (i) the central loop is big enough to accommodate the crystal, (ii) the resonant frequency of the LGR is close to 2.5\,GHz, and (iii) the resonator is overcoupled to the external circuitry. The LGR is capacitively coupled to an antenna pin, which is placed near the gap (see Fig.~\ref{fig:setup}(b)) to ensure strong extrinsic coupling. The extrinsic coupling can be improved by increasing the length of the pin and by moving the pin laterally closer to the gap. 
The reflection spectrum of the LGR as shown in Fig.~\ref{fig:setup}(b) is fitted to an element (eqn.~\ref{eqn:cavityReflection}) of the scattering matrix derived using input-output formalism for our system (see~\ref{app:theory}). 
We achieved the resonant frequency of 2625.3\,MHz (see Fig.~\ref{fig:setup}(c)) by iteratively removing layers from the gap using diamond abrasive paper. This also polished the gaps and improved the internal quality factor of the LGR. Details of resonator tuning after fabrication are presented in~\ref{app:cavCharac}. For optical coupling, the LGR has grooves on the front and rear walls to align the optics, which contain GRIN lenses and ferrules attached to fibers on each end, as shown in Fig.~\ref{fig:LGR_CAD}. The output fiber from the LGR is connected to a fast photodetector for heterodyne measurements. The detailed experimental setup is presented in~\ref{app:setup}.

\subsection{CW Transduction Protocol}

In Fig.~\ref{fig:setup}(d), the pulse sequence for CW transduction is shown, which consists of population initialization and transduction phases. The population initialization phase contains several optical pulses with wavelengths matched to different transitions. In each experimental cycle, the same base level of optical depth is initiated in the pump transition before the transduction phase. Following the initialization phase, the population is allowed to relax to the ground states. This also dissipates the local heat in the crystal generated by the strong optical pulses during the initialization. 

The transduction phase consists of a simultaneous drive of the pump and spin transitions. A 10\,ms long intense pump pulse drives the pump transition and establishes a steady-state population in the $\vert 1_e\rangle$ excited state. Simultaneously, a vector network analyzer (VNA) scans 30\,MHz around the spin transition with -45\,dBm power reaching the input port of the LGR. When the scan frequency matches the spin resonance, a transduced optical signal is generated. This signal is detected via heterodyne detection on the fast photodetector as shown in the setup in Fig.~\ref{fig:setup}(b). The heterodyne signal is amplified before being fed to the VNA for $S$-parameter measurements.

The leaked portion of the intense pump pulse is strong enough to drive the local oscillator for the heterodyne detection. It also provides a frequency reference for the transduced optical signal. Therefore, the beat frequency corresponds to the spin transition frequency ($\omega_t - \omega_p = \omega_m$). The $S_{21}$ plot on the VNA contains the information about efficiency across the frequency span. The efficiency is calculated by removing the complex offset ($S_{21}^{\mathrm{noise}}$, when the pump is OFF) from the signal ($S_{21}^{\mathrm{signal}}$, when the pump is ON) as described in~\ref{app:cmplx_trans_sgnl}.

\section{Results}
\subsection{Transduction Efficiency}\label{subsec:transEff}

\begin{figure}
\centering
\includegraphics[width=\linewidth]{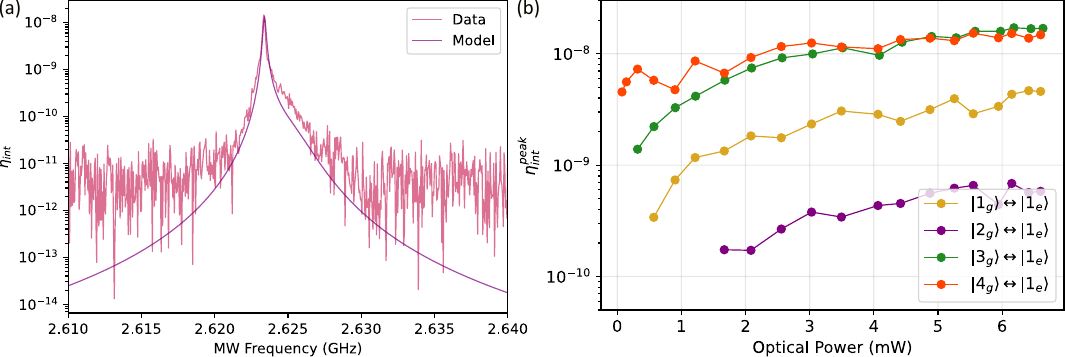}
\caption{(a) Experimental transduction efficiency spectrum fitted to our model with microwave frequency near the spin transition and $\vert4_g\rangle \leftrightarrow \vert1_e\rangle$ pump transition. The sharp peak corresponds to the spin transition, while the shoulder is due to the microwave cavity detuned by 1.94\,MHz from the spin transition. (b) Peak experimental efficiencies at varying optical powers for all the ground states to $\vert1_e\rangle$ pump transitions.}
\label{fig:eff_model_fit}
\end{figure}

In Fig.~\ref{fig:eff_model_fit}(a), we present the measured M2O transduction efficiency when the microwave field is scanned near the spin transition frequency, with $\vert 4_g \rangle \leftrightarrow \vert 1_e \rangle$ as the pump transition. The efficiency is extracted from the measured $S_{21}$ by accounting for microwave transmission losses, optical coupling losses, the leaked pump power and by calibrating the conversion gain of the fast photodetector. The experimental data is fitted using the theoretical model described in~\ref{app:theory}, with the corresponding parameters listed in Table~\ref{tab:parameters}.

The peak observed in Fig.~\ref{fig:eff_model_fit}(a) corresponds to the spin transition frequency, whereas the shoulder on the higher-frequency side arises from the LGR driving off-resonant transitions. A similar spectral feature has been reported with Yb:YVO coupled to a planar microwave resonator~\cite{xie-2025}. Figure~\ref{fig:eff_model_fit}(b) summarizes the peak transduction efficiencies obtained with different pump transitions as a function of pump power. The input microwave power is kept constant across all measurements involving the ground-state transitions. We find that the efficiencies associated with the $\vert 3_g \rangle$ and $\vert 4_g \rangle$ levels are the highest, primarily due to the stronger branching ratios of the corresponding optical transitions. Owing to the spectral isolation and experimental accessibility of the $\vert 4_g \rangle \leftrightarrow \vert 1_e \rangle$ transition, we identify it as the most suitable ground-state transition for subsequent transduction experiments.

\begin{figure}[t]
\centering
\includegraphics[width=\linewidth]{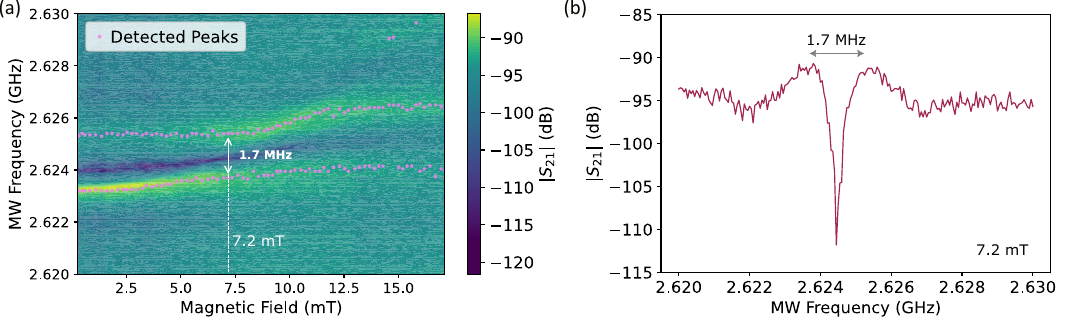}
\caption{(a) Heat map of transduction efficiency ($|S_{21}|$) with varying external magnetic field. Avoided crossing is clearly seen around 7.2\,mT. (b) Slice of 3D-plot in (a) at 7.2\,mT. After smoothening the curve, the peak separation is 1.7\,MHz.}
\label{fig:avoided_crossing}
\end{figure}

\subsection{Avoided Crossing / Spin-Cavity Coupling}

As discussed in~\cite{xie-2025}, the transduction efficiency depends on microwave ($C_m$), optical ($C_{o}$) and atomic cooperativities ($C_{atom}$), where cooperativity ($C = g_{ens}/\gamma_1\gamma_2$) is a ratio of
useful ensemble coupling strengths ($g_{ens}$) to non-useful decay couplings ($\gamma_i$) (see~\ref{app:theory}). To calculate $C_m$, we measure the microwave--spin ensemble coupling strength $g_{ens, m}$. When a two-level system (spin transition) interacts strongly with a resonant cavity, the eigenstates of the coupled system are no longer the bare cavity and spin states but hybridized light--matter states known as dressed states~\cite{kubo-2010}. In the strong-coupling regime, the dressed states exhibit an avoided crossing whose minimum splitting is $2g_{ens,m}$, giving direct access to $g_{ens,m}$.

To observe the dressed states, we bring the spin transition closer to the microwave cavity by varying the external magnetic field. Therefore, it is important that the spin transition curves towards the cavity with increasing magnetic field (see~\ref{app:crystalOrientation}). When the spin transition is close to the cavity, the hybridized states result in an avoided crossing as shown in Fig.~\ref{fig:avoided_crossing}(a). The minimum separation between the two peaks occurs at 7.2\,mT. 
As shown in Fig.~\ref{fig:avoided_crossing}(b), a slice plot at this magnetic field reveals the peak separation to be 1.7\,MHz, which gives the microwave ensemble coupling strength of $g_{m,ens}=2\pi \times 850$\,kHz. The microwave cooperativity derived in equation~\ref{eqn:C_m} at finite spin-cavity detuning is given by:

\begin{equation}
    C_m = \left|\frac{4g_{m,ens}^2}{\gamma_s (\kappa_{m}+2i\delta_{s,cav})} \right|
    \label{eqn:microwave_coop}
\end{equation}
where, $\delta_{s,cav}=2\pi \times 1.94$\,MHz is the spin-cavity detuning, $\kappa_{m}=2\pi \times 4$\,MHz is the cavity linewidth and $\gamma_s =2\pi \times  200$\,kHz is the spin inhomogeneous broadening extracted from fitting the model in Fig.~\ref{fig:eff_model_fit}(a). Feeding the parameters in eqn.~\ref{eqn:microwave_coop}, gives $C_m = 2.6$ (at 0 external magnetic field), implying that the system is in the strong cooperativity regime ($C_m >1$) for the spin-microwave interaction.

\subsection{Power characterizations}

\begin{figure}[h]
\centering
\includegraphics[width=\linewidth]{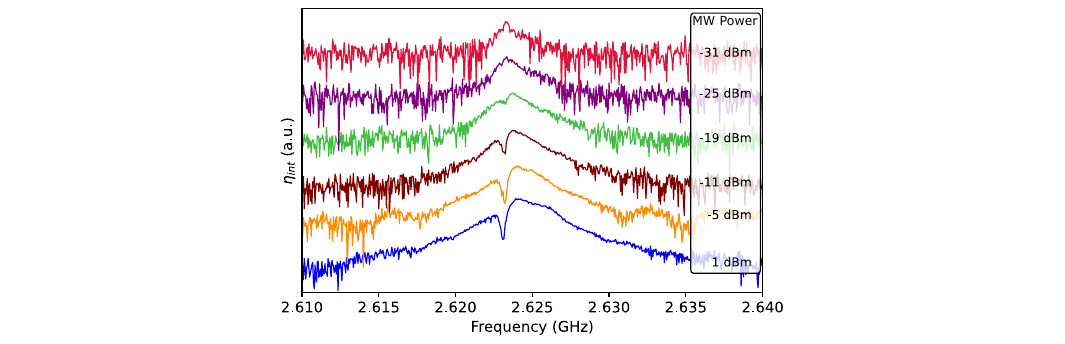}
\caption{Transduction efficiency spectrum with varying microwave power. The spin transition peak evolves into a dip at higher microwave powers due to saturation.}
\label{fig:power_char}
\end{figure}

To estimate the number of spins pumped into the excited state, we examine the microwave power required to saturate the spin transition. Accordingly, the microwave input power is swept from low to high values (-31--1\,dBm), and the corresponding transduction efficiency is shown in Fig.~\ref{fig:power_char}. At low microwave powers, the peak associated with the spin transition is clearly visible. As the microwave power is increased, this peak gradually evolves into a dip. 
Things can be understood as the power at which the number of microwave photons in the cavity becomes comparable to the number of spins available to absorb them. Beyond this point, additional microwave photons drive the spins back toward the first excited level $\vert1_e\rangle$, thereby reducing the transduction efficiency. This effect manifests as a dip in the measured efficiency. The microwave power at which this crossover occurs is identified as the saturation power.

From Fig.~\ref{fig:power_char}, the saturation power is estimated to lie between $-25\,\mathrm{dBm}$ and $-19\,\mathrm{dBm}$. During the measurement, the VNA scanned a spectrum of 1001 frequency points over a total duration of about $10\,\mathrm{ms}$, corresponding to an integration time of $10\,\mu\mathrm{s}$ per point. This implies that the total number of microwave photons incident on the cavity that saturates the spins is between $1.8\times10^{13}$ and $7.2\times10^{13}$.

We can calculate the intracavity photon number by the equation derived in~\ref{app:intracavity_Photons} as:
\begin{eqnarray}
    \frac{n_{cav}}{n_{in,e}} = \frac{\kappa_{m,e}}{\frac{\kappa_m^2}{4} + \delta_{m,cav}^2},
\end{eqnarray}
where, $n_{cav}$ and $n_{in,e}$ are the intracavity and input photon number of the cavity.
Of the total incident photons, only $\sim3.5\%$ enter the cavity at the spin resonance, resulting in an intracavity photon number between $6.3\times10^{11}$ and $2.5\times10^{12}$. Consequently, as long as the power is below the saturation power, the number of spins pumped into the excited state should match the intracavity photon numbers. The total number of spins within the optical mode volume is estimated to be $3.7\times 10^{14}$, indicating that only a fraction, $1.7\times10^{-3}$ to $6.7\times10^{-3}$ of the spins are excited. This experimental estimate of excited state population fraction is very close to the theoretically calculated value of $7.25\times10^{-3}$ using eqn.~\ref{eqn:n_2}.

\section{Discussion and Outlook}

In this work, we have characterized 5\,ppm doped Yb:YSO crystal for M2O transduction at zero applied magnetic field owing to the hyperfine splittings at GHz scale. With an LGR and single-pass optical configuration, we achieve internal CW transduction efficiency of $2\times 10^{-8}$ and a bandwidth of 200\,kHz. We also achieve spin-microwave cooperativity exceeding unity, demonstrated through avoided crossing measurements. Based on saturating microwave power, we estimate the number of spins in the interaction region close to the fitting parameters used to fit the experimental data to the model. The merits of using the long optical coherence and large optical inhomogeneous broadening of Yb:YSO for multiplexed memory-assisted M2O transducer is presented in ref.~\cite{gautam-2026}.

The large optical and spin inhomogeneous linewidths in Yb:YSO result in small cooperativities (eqns~\ref{eqn:C_m},~\ref{eqn:C_o},~\ref{eqn:C_atom}) which causes reduced transduction efficiency. The large mode volume provided by LGR enhances the microwave interaction, but also limits the pump Rabi frequency, which further reduces the efficiency via reduced atomic cooperativity (eqn~\ref{eqn:C_atom}) and reduced steady-state excited state population ($n_2$, see eqn.~\ref{eqn:n_2}). The doping concentration also affects the transduction efficiency due to weak ensemble coupling strengths. Finally, narrow optical homogeneous linewidth (or longer optical coherence time) also causes reduced efficiency through the reduced steady-state population in the excited state ($n_2$, see eqn.~\ref{eqn:n_2}).

To improve the efficiency of our system, we have a few factors to improve up on. An impedance matched optical cavity, resonant to the transduced signal can improve the efficiency by up to $3.5\times$. A small beam diameter ($150\,\mu$m instead of currently employed $500\,\mu$m) can improve the transduction efficiency by up to 2 orders of magnitude. However, this will increase the difficulty in optical alignment and possibly reduced overall optical coupling. We can increase the peak pump power to 10\,mW to achieve $3\times$ better efficiencies. The spin-cavity detuning can be further reduced (from 1.9\,MHz to $<1$\,MHz) to improve the efficiency by more than $1.5\times$. Improving the external microwave coupling (from 0.425 to 0.85) of the LGR can boost the efficiency by $2\times$. Finally, using a higher doping concentration crystal 10\,ppm or 50\,ppm Yb:YSO crystal can boost the efficiency by up to 2 to $7\times$. Overall, the internal transduction efficiency can be improved to $\sim 10^{-4}$. The enhancement factors of various parameters are discussed in~\ref{app:enhancement}.

\section{Acknowledgments}
We acknowledge S. Barzanjeh for helpful discussions. We thank A. Mohamed for help with the microwave switches and transmission lines in the dilution fridge and A. El-Hamamsy for help with the VNA.

\section*{References}
\bibliography{References}

\clearpage

\appendix
\section{Theory}\label{app:theory}
\subsection{Input-output Formalism}\label{app:in_out_formalism}
The experimental results presented in the main paper can be modeled by a three-level system interacting with an optical and a microwave cavity as described in ref.~\cite{xie-2025}. Since, a single-atom picture can very closely resemble that of a multi-atom picture as mentioned in the ref.~\cite{xie-2025}, we stick to the single-atom picture to describe our system. Our goal here is to find tuning parameters to improve the efficiency of our system.

\begin{figure}[h]
    \centering
    \includegraphics[width=0.5\linewidth]{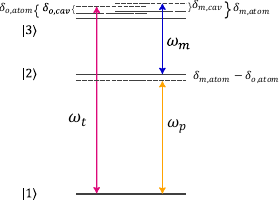}
    \caption{\textbf{Level Diagram.} A three-level V-type system interacting with optical and microwave fields. Transition $\vert1\rangle \leftrightarrow \vert2\rangle$ is driven by an optical pump at frequency $\omega_p$. Transitions $\vert1\rangle \leftrightarrow \vert3\rangle$ and $\vert2\rangle \leftrightarrow \vert3\rangle$ interact with an optical cavity and driven by a microwave cavity respectively. Detunings of cavity and atomic transitions from the fields are shown.}
    \label{fig:placeholder}
\end{figure}

We have a three-level system, interacting with a single-port microwave cavity and a single-pass optical configuration. The single-pass optical configuration can be approximated as a very weak optical cavity formed due to reflection from the surfaces of the crystal. A more complete treatment would instead solve the coupled atom-field equations as a function of propagation distance $z$ through the crystal, tracking the buildup of the optical field amplitude and the local atomic coherence together as the field traverses the ensemble, rather than assuming a single, spatially uniform intracavity mode. Such a propagation-dependent (Maxwell-Bloch-type) treatment would capture effects such as reabsorption and the spatial dependence of the atom-field coupling along the beam path, which the weak-cavity approximation neglects. We adopt the single-mode weak-cavity approximation here for analytical tractability, and its validity is supported by the good agreement between the resulting model and the experimental data (Figure~\ref{fig:eff_model_fit}). It also allows us to estimate the effect of incorporating a high-finesse optical cavity for optimal efficiency.

In the current setup, the optical field is collected and delivered through a single-mode fiber, which supports only the fundamental spatial mode (LP$_{01}$) at the operating wavelength. Because coupling into and out of the fiber inherently projects the field onto this single spatial mode, the optical field can be treated as a single bosonic mode $\hat{a}$ despite the finite transverse extent of the atomic ensemble in the crystal; atoms outside the effective mode volume defined by the fiber-coupled beam waist do not contribute appreciably to the coherent interaction and are accounted for through the ensemble coupling strength introduced below, rather than through additional field modes. The system Hamiltonian in the frame rotating with respect to the field frequencies can be written as:

\begin{eqnarray}
    H/\hbar  = &\delta_{o,cav}\hat a^\dagger \hat a + \delta_{m,cav}\hat b^\dagger \hat b + (\delta_{m,atom}-\delta_{o,atom})\hat \sigma_{22} -\delta_{o,atom}\sigma_{33} \\ \nonumber
    & + g_{o,ens}(a^\dagger \hat \sigma_{13} + h.c.) + g_{m,ens}(b^\dagger \hat \sigma_{23} + h.c.) + \Omega_{p}(\hat \sigma_{12} + h.c.) 
\end{eqnarray}
where $\hat a$ and $\hat b$ are the operators for the optical and microwave modes respectively, $\delta_{o/m, cav}$ are the field-cavity detunings, $\delta_{o/m, atom}$ are the field-atom detunings, $\Omega_p$ is the pump Rabi frequency, and $\sigma_{ij}$ are the atomic transition operators for $i,j\in \{1,2,3\}$. Here $g_{o, ens}$ and $g_{m,ens}$ denote the \emph{ensemble} (collective) optical and microwave coupling strengths rather than single-atom couplings; they are related to the corresponding single-atom couplings $g_{o}$ and $g_{m}$ by $g_{o,ens} = \sqrt{N_{o}}\,g_{o}$ and $g_{m,ens} = \sqrt{N_{m}}\,g_{m}$, where $N_{o}$ and $N_{m}$ are the effective numbers of ions coherently coupled to the optical and microwave modes respectively. This collective enhancement follows from the coherent addition of the individual atomic dipoles within the mode volume and is what allows the single-atom-like equations of motion below to describe the full ensemble response.

Assuming the operators to be complex numbers, using the input-output formalism and neglecting fast rotating terms, we get:

\begin{eqnarray}
    \dot a &= -\frac{i}{\hbar} [a,H] -\frac{\kappa_o}{2} a + \sqrt{\kappa_{o,e,2}}a_{in, e,1} + \sqrt{\kappa_{o,i,1}}a_{in, i,1} \\ \nonumber
    &\ \ \ \ \ \ \ \ \ \ \ \ \ \ \ \ \ \ \ \ \ \ \ \ \ + \sqrt{\kappa_{o,e,2}}a_{in, e,2} + \sqrt{\kappa_{o,i,2}}a_{in, i,2} \\
    \dot \sigma_{13} &= -\frac{i}{\hbar} [\sigma_{13},H] -\frac{\gamma_o}{2} \sigma_{13} \\
    \dot \sigma_{23} &= -\frac{i}{\hbar} [\sigma_{23},H] -\frac{\gamma_s}{2} \sigma_{23} \\
    \dot b &= -\frac{i}{\hbar} [b,H] -\frac{\kappa_m}{2} b + \sqrt{\kappa_{m,e}}b_{in, e} + \sqrt{\kappa_{m,i}}b_{in, i}\label{eqn:inout_mw}
\end{eqnarray}
where $\kappa$ are the total cavity decay rates, $\kappa_{x,e/i}$ are the extrinsic/intrinsic decay rates from ports 1 or 2, while the terms $a/b_{in/out,e}$ are the input/output field modes from/to the external environment and $a/b_{in/out,i}$ are the input/output field modes due to internal losses. $\gamma_{o}$ and $\gamma_{s}$ are single atom decay terms for the optical and spin coherences respectively; in the single-atom picture used here, they are approximated by the respective inhomogeneous linewidths of the ensemble, which are the quantities directly accessible from spectroscopic measurements.

We can collect all the operators in vectors as:
\begin{eqnarray}
    X &= [a, \sigma_{13}, \sigma_{23},b]^T \\
    X_{in} &= [a_{in,e,1}, a_{in,i,1},a_{in,e,2},a_{in,i,2}, 0, 0, b_{in,e}, b_{in,i}]^T \\
    X_{out} &= [a_{out,e,1}, a_{out,i,1}, a_{out,e,2}, a_{out,i,2}, 0, 0, b_{out,e}, b_{out,i}]^T
\end{eqnarray}

Now, writing the above differential equations in terms of these vectors in the frequency domain, we get:

\begin{eqnarray}
    \dot X &= 0 = A X + B X_{in} \\
    X_{out} &= B^T X - X_{in} \\
    X_{out} &= SX_{in}
\end{eqnarray}

This gives the scattering matrix $S$ in terms of the matrices $A$ and $B$.

\begin{equation}
    S = B^T[-A]^{-1}B - I_8
\end{equation}
where, $I_k$ is an identity matrix of dimension $k$.

We can write the matrices $A$ and $B$ as:
\begin{equation}
\fl    A = \left[
\begin{array}{cccc}
-i\delta_{o,cav}-\frac{\kappa_o}{2} & -ig_{o} & 0 & 0 \\
-ig_{o}(n_1-n_2) & -i\delta_{o,atom}-\frac{\gamma_o}{2} & i\Omega_p & 0\\
0 & i\Omega_p & -i\delta_{m,atom}-\frac{\gamma_{spin}}{2} & -ig_{m}(n_2-n_3)\\
0 & 0 & -ig_{m} & -i\delta_{m,cav}-\frac{\kappa_m}{2}
\end{array}
\right]
\end{equation}

\begin{equation}
\fl    B = \left[
\begin{array}{cccccccc}
\sqrt{\kappa_{o,e,1}} & \sqrt{\kappa_{o,i,1}} & \sqrt{\kappa_{o,e,2}} & \sqrt{\kappa_{o,i,2}} & 0 & 0 & 0 & 0\\
0 & 0 & 0 & 0 & 0 & 0 & 0 & 0\\
0 & 0 & 0 & 0 & 0 & 0 & 0 & 0\\
0 & 0 & 0 & 0 & 0 & 0 & \sqrt{\kappa_{m,e}} & \sqrt{\kappa_{m,i}}
\end{array}
\right]
\end{equation}
where, $n_i$ is the fraction of population in level $i$ for $i \in \{1,2,3\}$.
Solving for $S$ and extracting the M2O transduction efficiency term $\eta_{M2O}= |S_{1,7}|^2$. At zero detunings, collecting the terms in a similar manner as ref.~\cite{xie-2025}, we get:

\begin{eqnarray}
    \eta_{M2O} = |S_{1,7}|^2 &= \frac{\kappa_{o,e,2}}{\kappa_o} \frac{\kappa_{m,e}}{\kappa_m} \frac{n_2-n_3}{n_1-n_3} \frac{4C_{m}C_oC_{atom}}{[(1+C_m)(1+C_o)+C_{atom}]^2}, \label{eqn:eta}
\end{eqnarray}
where, the cooperativity terms at zero detunings are:

\begin{eqnarray}
    C_m &= \frac{4g_m^2(n_2-n_3)}{\kappa_m \gamma_s}\label{eqn:C_m}, \\
    C_o &= \frac{4g_o^2(n_1-n_3)}{\kappa_o \gamma_o} \label{eqn:C_o}, \\
    C_{atom} &= \frac{4\Omega_p^2}{\gamma_o \gamma_s}\label{eqn:C_atom}.
\end{eqnarray}

\noindent We can also fit the cavity reflection spectrum using the term $S_{7,7}$, which gives us:
\begin{eqnarray}\label{eqn:cavityReflection}
    |S_{7,7}| = \left| 1- \frac{2\kappa_{m,e}}{(\kappa_{m}+2i\delta_{m,cav})} \right|.
\end{eqnarray}

\subsection{Microwave Photons in the Cavity}\label{app:intracavity_Photons}
The microwave cavity port is a boundary where microwave photons partially enter into the cavity while the remaining photons reflects based on the impedance matching condition. The fraction of photons that enters into the cavity can be derived using eqn.~\ref{eqn:inout_mw} by assuming that the Hamiltonian only consists of the microwave cavity terms. In the reference frame of the microwave field frequency and in the Fourier domain, the equation can be written as:

\begin{eqnarray}
    0=-i\delta_{m,cav}b - \frac{\kappa_m}{2}b + \sqrt{\kappa_{m,e}}b_{in, e} + \sqrt{\kappa_{m,i}}b_{in, i},
\end{eqnarray}
which result in the relation between the internal field amplitude, $b$ and the input field amplitude, $b_{in,e}$ as:

\begin{eqnarray}
    b = \frac{\sqrt{\kappa_{m,e}}b_{in, e} + \sqrt{\kappa_{m,i}}b_{in, i}}{\frac{\kappa_m}{2} + i\delta_{m,cav}},
\end{eqnarray}
which can be further simplified, assuming the intrinsic field amplitude, $b_{in,i}$ is negligible. The fraction of photons entering the cavity at detuning $\delta_{m,cav}$ from the cavity resonance is given by:

\begin{eqnarray}
    \frac{n_{cav}}{n_{in,e}} = \frac{\kappa_{m,e}}{\frac{\kappa_m^2}{4} + \delta_{m,cav}^2},\label{eqn:photon_frac}
\end{eqnarray}
where, $n_{cav}$ is the number of microwave photons inside the cavity and $n_{in,e}$ is the number of microwave photons sent at the input port of the cavity.

\subsection{Parameters}\label{app:parameters}

\begin{table}[htbp]
    \centering
    \begin{tabular}{|l|c|c|}
    \hline
        Parameter & Symbol & Value \\ \hline 
        LGR Total Linewidth  & $\kappa_m$ & $2\pi\times 4$ MHz\\
        LGR Q-factor & $Q$ & 653 \\
        LGR internal loss rate & $\kappa_{m,i}$ & $2\pi\times2.3$  MHz\\
        LGR external loss rate & $\kappa_{m,e}$ & $2\pi\times1.7$\,MHz\\
        Spin-cavity coupling strength & $g_{m,ens}$ & $2\pi\times 850$\,kHz \\
        Spin-cavity detuning & $\delta_{s,cav}$ & $2\pi\times1.94$\,MHz \\
        Spin inhomogeneous broadening & $\Gamma_{s}$ & $2\pi\times 200$\,kHz$^*$ \\
        Optical inhomogeneous broadening & $\Gamma_{o}$ & $2\pi\times 390$\,MHz  \\
        Optical Coherence Time & $T_{2,o}$ & $800\,\mu$s  \\
        Population lifetime & $T_{1}$ & $1.3\,$ms  \\
        Optical coupling strength & $g_{o,ens}$ & $2\pi\times 1.3$\,GHz \\
        Pump Rabi frequency & $\Omega_{p}$ & $2\pi\times200$\,kHz \\
        Optical Cavity Finesse & $\mathcal{F}$ & 0.98$^\dagger$ \\
        Optical Free-spectral Range & $FSR$ &  $2\pi\times8.3$\,GHz$^\dagger$ \\
        Microwave Cooperativity & $C_m$ & 2.6  \\ 
        Optical Cooperativity & $C_o$ & $8.4\times 10^{-2}$$^\dagger$ \\
        Atomic Cooperativity & $C_{atom}$ & $0.18$$^\dagger$ \\
        Population fraction - ground state & $n_1$ & $\sim 1$  \\
        Population fraction - first excited state & $n_2$ & $7.25\times 10^{-3}$$^\dagger$ \\
        Population fraction - second excited state & $n_3$ & $0$ \\
        Calculated Efficiency & $\eta_{cal}$ & $2.8\times10^{-8}$ \\
        Experimental Efficiency & $\eta_{exp}$ & $1.4\times10^{-8}$ \\ \hline
    \end{tabular}
    \caption{Parameters used to fit the experimental data with the model in Fig.~\ref{fig:eff_model_fit}. $^\dagger$Estimated values. $^*$Fitted values.}
    \label{tab:parameters}
\end{table}

The theoretical model developed in section~\ref{app:in_out_formalism} requires various parameters to match the experimental results in Fig.~\ref{fig:eff_model_fit}. These parameters are listed out in Table~\ref{tab:parameters}. Most of the values are measured or extracted from experimental data, while some of them are estimates. The value of spin inhomogeneous broadening is set to 200\,kHz so that the peak from the model resembles that of the experimental data. The following discussion describes how each of the parameter values are extracted.

\begin{figure}
    \centering
    \includegraphics[width=0.6\linewidth]{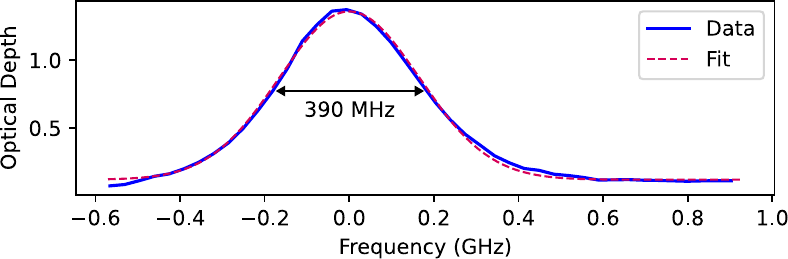}
    \caption{\textbf{Optical Inhomogeneous Broadening.} $\vert4_g\rangle \leftrightarrow \vert1_e\rangle$ optical inhomogeneous broadening fitted to Gaussian function.}
    \label{fig:4g_1e_optical_depth_fit}
\end{figure}

The LGR cavity parameters are extracted by fitting cavity reflection spectrum in Fig.~\ref{fig:setup}(b) using equation~\ref{eqn:cavityReflection}. Spin-cavity coupling strength is measured from the peak separation in the avoided crossing plot in Fig.~\ref{fig:avoided_crossing}(a). Spin-cavity detuning is calculated from the LGR resonance in Fig.~\ref{fig:setup}(b) and spin transition peak in the M2O transduction spectrum Fig.~\ref{fig:eff_model_fit}(a). Optical inhomogeneous broadening is measured from the Gaussian fitting of optical spectrum of $\vert4_g\rangle \leftrightarrow \vert1_e\rangle$ transition as shown in Fig.~\ref{fig:4g_1e_optical_depth_fit}. Optical coherence time is taken from~\cite{gautam-2026} and the population lifetime is taken from~\cite{welinski-2016}.
Pump Rabi frequency is measured via 2-Pulse photon echo measurements on the $\vert4_g\rangle \leftrightarrow \vert1_e\rangle$ transition. 

For optical cavity finesse $\mathcal{F}$, we use the YSO crystal's refractive index ($n_c=1.8$) which results in reflectivity of $R=(n_c-1)^2/(n_c+1)^2\sim 8\%$ at each end surface. Finesse is then calculated by $\mathcal{F}=\pi\sqrt{R}/(1-R)=0.977$. Free-spectral range is calculated using $FSR = c/2n_cL= 2\pi\times 8.3\,$GHz, where the length of the crystal is 10\,mm. The optical cavity linewidth is calculated using $\kappa_o = FSR/\mathcal{F} =2\pi\times 8.6\,$GHz.

To calculate optical cooperativity $C_o$ using eqn~\ref{eqn:C_o}, we need optical ensemble coupling constant $g_o$, which is calculated using:
\begin{equation}\label{eqn:optical_coup_strength}
    g_o = d \sqrt{\frac{\omega N}{2\hbar\epsilon_0 n_c^2 V_m}}
\end{equation}
where, $d$ is the dipole moment of the $\vert2_e\rangle \leftrightarrow \vert4_g\rangle$ transition, $N/V_m$ is the density of atoms in the crystal, and $\omega = 2\pi c/\lambda$ for $\lambda = 978.8\,$nm. The dipole moment $d$ is calculated using~\cite{scully-1997}:
\begin{equation}
    d = \sqrt{\frac{3\pi \epsilon_0 \hbar c^3}{\omega^3 \chi_{\textrm{local}} T_1}}
\end{equation}
where $T_1$ is the population lifetime of the excited level in the two-level system $\vert 2_e\rangle \rightarrow \vert4_g\rangle$ and $\chi_{\textrm{local}}=n(n^2+2)^2/9$ is the local field correction factor~\cite{crenshaw-2000}. Since, the excited state lifetime of $\vert2_e\rangle$ is $T_1(\vert2_e\rangle)=1.3\,$ms, the decay lifetime from $\vert2_e\rangle \rightarrow \vert4_g\rangle$ can be calculated using the branching ratio as $T_1(\vert2_e\rangle \rightarrow \vert4_g\rangle)=1.3/ 0.04= 32.5\,$ms. The value of dipole moment for $\vert2_e\rangle \leftrightarrow \vert4_g\rangle$ transition is therefore, $d=1.4\times 10^{-32}\,$C\,m. 


For estimation of the density of atoms $N/V_m$, we use the doping concentration of the crystal (5\,ppm) to get $4.7\times 10^{22}$ atoms/m$^3$. The value of optical ensemble coupling strength is calculated to be $g_o = 2\pi \times 1.3\,$GHz and the optical cooperativity is $C_o = 1.45$. 
Atomic cooperativity $C_{atom}$ is calculated from eqn.~\ref{eqn:C_atom} using the values of the optical and spin inhomogeneous broadenings. 

The population fraction is estimated from steady-state condition of an optically-driven two-level system~\cite{steck-2007}:
\begin{eqnarray}
    \rho_{ee}(t\longrightarrow \infty) = \frac{\Omega^2}{2\gamma_{\mathrm{h}}\Gamma} \frac{1}{1+\frac{\Delta^2}{\gamma_{\mathrm{h}}^2} + \frac{\Omega^2}{\gamma_{\mathrm{h}}\Gamma}},
    \label{eqn:n_2}
\end{eqnarray}
where $\gamma_{\mathrm{h}} = 1/T_{2,o}$, is the homogeneous linewidth and $\Gamma = 1/T_1$, is the population decay rate from the excited to the ground state. Integrating over the optical inhomogeneous broadening gives the steady-state population fraction in the excited state.

\subsection{Enhancement Factors}\label{app:enhancement}
\begin{figure}[h]
    \centering
    \includegraphics[width=\linewidth]{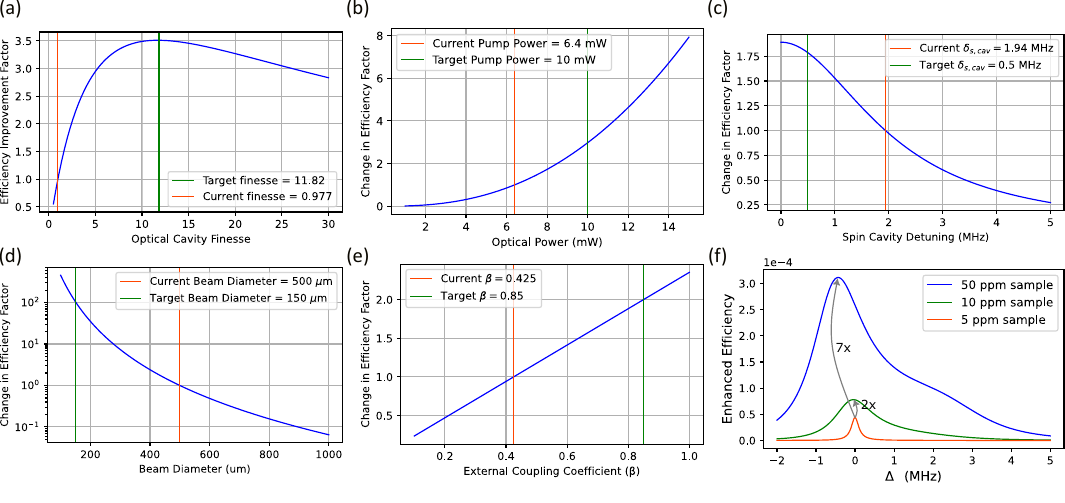}
    \caption{\textbf{Efficiency enhancement.} Improvement in efficiency due to different tuning parameters. The final efficiencies with target parameters for two different doping concentration is shown in (f).}
    \label{fig:enhancementFactor}
\end{figure}

The enhancement factors presented in Fig.~\ref{fig:enhancementFactor} are calculated by varying one parameter at a time while keeping all the other parameters fixed at their current experimental values listed in Table~\ref{tab:parameters}. For each parameter, the transduction efficiency is computed using eqn.~\ref{eqn:eta} and normalized to the current experimental efficiency to obtain the enhancement factor. The combined target efficiency shown in Fig.~\ref{fig:enhancementFactor}(f) is obtained by updating all the parameters to their target values and evaluating the resulting efficiency for 5, 10 and 50\,ppm doping concentrations. For each doping concentration, their respective parameters like $T_{2,o}$, $\Gamma_{o,s}$, etc. are used. The efficiency significantly depends on the coherence time $T_{2,o}$ --- with better efficiencies for lower coherence time. 

Pushing the efficiency beyond $10^{-4}$ therefore requires more than optimizing a single parameter in isolation, since the terms entering eqn.~\ref{eqn:eta} are strongly interdependent. A shorter optical coherence time improves efficiency by allowing more population to be initialized in the excited state (eqn.~\ref{eqn:n_2}) and therefore raising $C_o$, but this benefit is only realized if the inhomogeneous linewidth remains narrow and the doping concentration is simultaneously increased to boost the ensemble coupling strength; higher doping, however, tends to broaden both the inhomogeneous linewidth and the homogeneous linewidth through cross-relaxation and superhyperfine interactions, which in turn works against the efficiency gain. An ideal material for high-efficiency CW transduction must therefore combine a short coherence time, a narrow inhomogeneous linewidth, and a high doping concentration without significant broadening---a figure of merit that few rare-earth-doped crystals satisfy simultaneously.

Reference~\cite{xie-2025} has achieved percent-level efficiency using Yb:YVO, whose very short optical and spin coherence times make it well suited to this regime, and this efficiency could plausibly be pushed further using the same approach. However, choosing a crystal with such short coherence times defeats the purpose of integrating a quantum memory with the transducer, as presented in the next chapter. For memory-assisted transduction, the figure of merit is different: the ideal material must instead have a sufficiently long coherence time to support synchronization at BSM nodes, while still maintaining narrow inhomogeneous broadening and high doping concentration.

\clearpage

\section{Experimental Setup}\label{app:setup}

\begin{figure}[htbp]
    \centering
    \includegraphics[width=\linewidth]{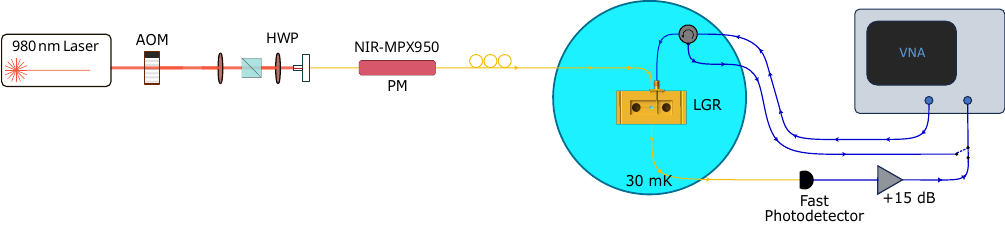}
    \caption{\textbf{Experimental setup.} Optical field is driven by a 980\,nm laser and pulsed with an accousto-optic modulator. The optical field is polarized before entering a phase modulator for efficient phase shifting. The phase modulator shifts the laser frequency to drive different optical transitions. The optical field is polarized to the $\vec{D_2}$ crystal axis before entering the fridge. The output optical field is detected on a fast-photodetector. A vector network analyzer (VNA) scans the LGR spectrum. The VNA also drives the chirped microwave pulses and records the amplified fast-photodetector response for transduction spectrum measurements.}
    \label{fig:detail_setup}
\end{figure}

The experimental setup consists of the microwave, optical and the dilution fridge as presented in Fig.~\ref{fig:detail_setup}. A free-running Toptica DL Pro laser is parked at the pump transition e.g. $\vert 4_g\rangle \rightarrow \vert 1_e\rangle$. The laser beam passes through an acousto-optic modulator (AOM) and a phase modulator (PM) before entering the dilution fridge. AOM (Model: AA Opto-Electronic: MT200-B100A0,5-1064) is used to create optical pulses. PM (Model: iXblue: NIR-MPX950) is used to shift the frequencies ($\sim \rm{GHz}$) for optical pumping and pump pulses. Two polarization controllers (half-wave plates) are placed before and after the PM for efficient frequency shift and for aligning the polarization along the $\vec{D}_2$-crystal axis to achieve maximum absorption~\cite{tiranov-2018}, respectively. Optical coupling efficiency through the crystal, including fiber and splice losses, is about $10 \%$. The output fiber is coupled to a fast photo-detector (Model: New Focus 1554-B).

On the microwave side, the LGR is driven by a VNA (Keysight-E5080B). To characterize the microwave cavity, the reflection port from the circulator is fed to the VNA to plot $S_{21}$-curve showing reflection spectrum of the LGR. To observe transduction, the output of the fast photo-detector is amplified and fed to the VNA. The resulting $S_{21}$-curve represents transduction signal at the corresponding input microwave frequency (\ref{app:cmplx_trans_sgnl}).

The losses in the microwave line inside the dilution fridge is estimated using two cryogenic microwave switches placed inside the fridge. There are four microwave lines inside the fridge, so we measured four combinations of losses i.e. $L_1 + L_2,\ L_1 + L_3, \ L_1 + L_4$ and $L_2 + L_4$. The approximation involved neglecting the losses in the connections between switches and circulator, which typically has a loss of about 0.2~dB.

The conversion gain of the fast photodetector is calibrated in DC mode, following the procedure recommended in its datasheet. The detector is illuminated with a known input optical power, and the corresponding voltage output is recorded on an oscilloscope. In practice, this involves building a table of three columns---AWG amplitude, optical power measured on the power meter, and the corresponding oscilloscope voltage reading---and applying a linear fit to the optical power versus voltage data to obtain the detector's conversion gain.

\section{Complex Transduced Signal}\label{app:cmplx_trans_sgnl}

\begin{figure}[htbp]
\centering
\includegraphics[width=\linewidth]{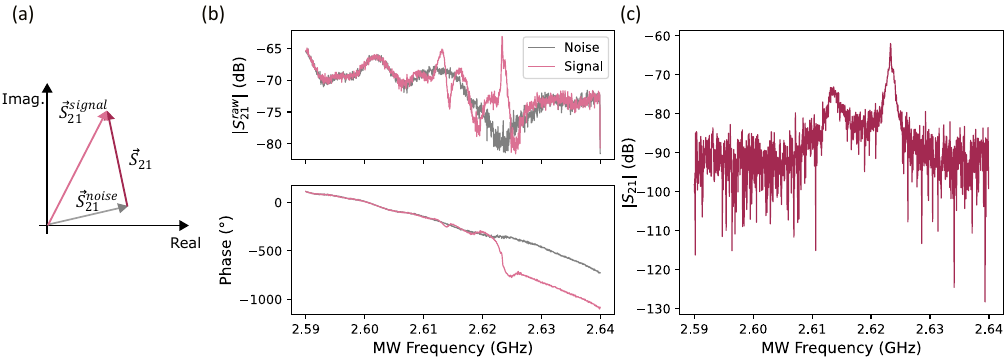}
\caption{(a) Complex $S_{21}$ vectors showing noise ($\vec{S}_{21}^{\mathrm{noise}}$), detected signal ($\vec{S}_{21}^{\mathrm{signal}}$) and transduced signal ($\vec{S}_{21}$). (b) Magnitude (top) and phase (bottom) of $\vec{S}_{21}^{\mathrm{raw}}$ which includes $\vec{S}_{21}^{\mathrm{noise}}$ for pump OFF case and $\vec{S}_{21}^{\mathrm{signal}}$ for pump ON case. (c) Magnitude of noise de-embedded complex vector $\vec{S}_{21}$ across the spin transition. The narrower peak on the right is the spin transition while the peak on the left is due to LGR at 2613.7\,MHz.}
\label{fig:M2O_typical}
\end{figure}

The VNA measures complex signals for the scattering parameters which means the signals have amplitude and phase components to them. In our experiments, when there is no pump going into the system, the VNA measures the complex noise, but when the pump drives the transition and transduction signal is generated, the complex signal is added to the complex noise floor. Therefore, to de-embed noise in such signal, a complex difference of the signal and the noise should be performed. A vector analysis of the operation is shown in Fig.~\ref{fig:M2O_typical}(a). 

An example of data analysis is shown in Fig.~\ref{fig:M2O_typical}(b). The top (bottom) plot contains the amplitude (phase) information when the pump is OFF (grey) vs ON (pink). Note that the direct measurement of the amplitude does not explicitly show the transduction peaks. After de-embedding the noise from the signal, the amplitude is plotted as shown in Fig.~\ref{fig:M2O_typical}(c). The data shown in Fig.~\ref{fig:M2O_typical}(b) and (c) corresponds to the cavity resonance of 2613.7\,MHz which is different than the data presented in Fig.~\ref{fig:eff_model_fit}, where the cavity resonance is 2625.3\,MHz.

\clearpage

\section{Crystal Orientation}\label{app:crystalOrientation}

\begin{figure}
    \centering
    \includegraphics[width=\linewidth]{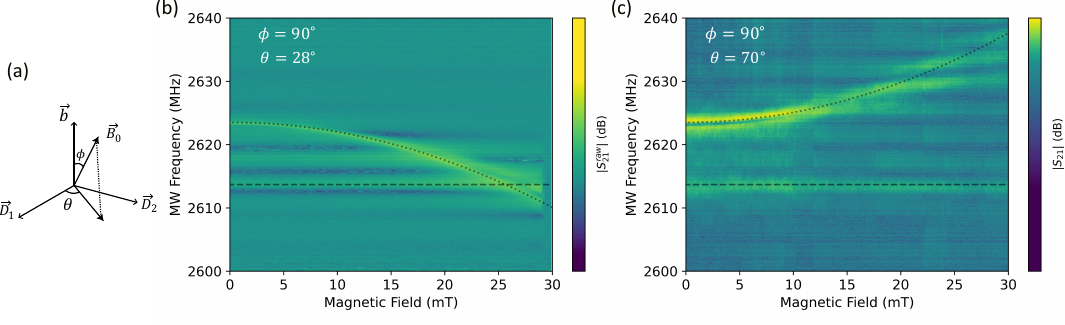}
    \caption{(a) Orientation of magnetic field in the frame of the crystal in spherical coordinates. (b) Heat map of measured $\vec{S}_{21}^{signal}$ with magnetic field. The spin transition (dotted line) curves towards the LGR resonance (dashed line) at 2613.7\,MHz. (c) Heat map of noise-dembedded $\vec{S}_{21}$ with magnetic field. The spin transition curves away from the LGR resonance at 2613.7\,MHz.}
    \label{fig:spinCurvature}
\end{figure}

Here, we calculate the curvature of the spin transition with the magnetic field. In the frame of the crystal, let $\vec{B}$ be aligned at an angle $\phi$ to the $\vec{b}$-axis and its projection on the $\vec{D_1}-\vec{D_2}$ plane makes an angle $\theta$ from the $\vec{D_1}$-axis (see Fig.~\ref{fig:spinCurvature}(a)). In the crystal frame of reference the Hamiltonian for Yb:YSO states are given by:
\begin{equation}
    H = \textbf{I}\cdot\textbf{A}\cdot\textbf{S} +\mu_B \textbf{B}\cdot \textbf{g}\cdot \textbf{S}
\end{equation}
where, \textbf{I} is the nuclear spin, \textbf{S} is the electronic spin, \textbf{A} is the hyperfine interaction tensor, \textbf{g} is the electronic Zeeman interaction tensor and $\mu_B$ is the electronic Bohr magneton. Here, we have ignored the nuclear Zeeman interaction term and used the values of \textbf{A} and \textbf{g} from ref.~\cite{tiranov-2018}. 

The energy of any individual state at a given magnetic field is calculated by the eigen-energy of the corresponding Hamiltonian. Therefore, the energy of $\vert1_e\rangle$ and $\vert2_e\rangle$ is the lowest and the second lowest eigen-energy of the excited state Hamiltonian with some offset. The spin curvature that we observe in Fig.~\ref{fig:spinCurvature}(b) and (c) is given by $E(\vert2_e\rangle)-E(\vert1_e\rangle)+E_0$, where $E_0$ is an offset. Here, we assume that the pump is always resonant to the $\vert4_g\rangle \leftrightarrow \vert1_e\rangle$ transition at small magnetic fields.

The spin curvature curves upwards or downwards depending on the angles $\theta$ and $\phi$. For the data in Fig.~\ref{fig:spinCurvature}(b) and (c), the best fit angles are shown on the plot. Therefore, we can adjust the magnetic field orientation with respect to the crystal in order to bring the spins closer to the cavity to observe avoided crossing or dressed state formation.

\clearpage
\section{Loop-gap resonator}\label{app:LGR_sim}
\subsection{Design}

\begin{figure}[h]
    \centering
    \includegraphics[width=\linewidth]{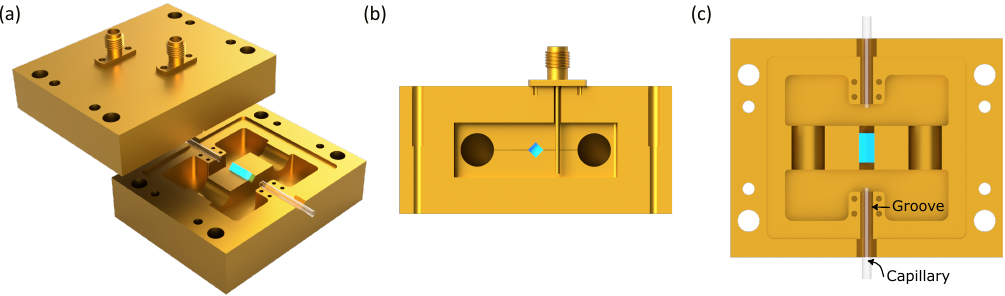}
    \caption{\textbf{LGR CAD Rendering.} (a) Open view of the LGR with input and output microwave ports. (b) Cross-section of the LGR showing the loops and the gaps. Capacitively coupled microwave antenna pin is also shown. (c) Top view of the bottom half of the LGR showing the placement of the crystal and capillaries along the groove. The capillaries are held by thin copper strips (not shown) screwed in the holes shown around the capillaries. These screws are used to align the optical coupling through the crystal. The flat area besides the crystal are the surface of the gaps. The semi-cylindrical surfaces beside the gap surfaces correspond to the loops.}
    \label{fig:LGR_CAD}
\end{figure}

The CAD rendering of the LGR is shown in Fig.~\ref{fig:LGR_CAD}. The resonator body is machined from a single block of copper and splits into two halves along the horizontal plane for crystal loading and optical alignment. The two gaps are formed by the flat surfaces visible on either side of the central diamond loop (Fig.~\ref{fig:LGR_CAD}(c)), and their widths and gaps, directly set the capacitance that determines the resonant frequency. The antenna pin enters from the side and terminates near one of the gaps to achieve capacitive coupling (Fig.~\ref{fig:LGR_CAD}(b)). As shown in Fig.~\ref{fig:LGR_CAD}(c), the capillaries rest in grooves that run through the front and rear walls of the resonator, and are secured by thin copper strips fastened with screws positioned around the capillaries. These screws provide mechanical degrees of freedom for aligning the optical path through the crystal. The overall dimensions of the resonator are constrained by the requirement to fit within the available space on the dilution fridge base plate while maintaining sufficient thermal contact for operation at 30 mK.

\subsection{Characterization}
\label{app:cavCharac}
\begin{figure}
    \centering
    \includegraphics[width=\linewidth]{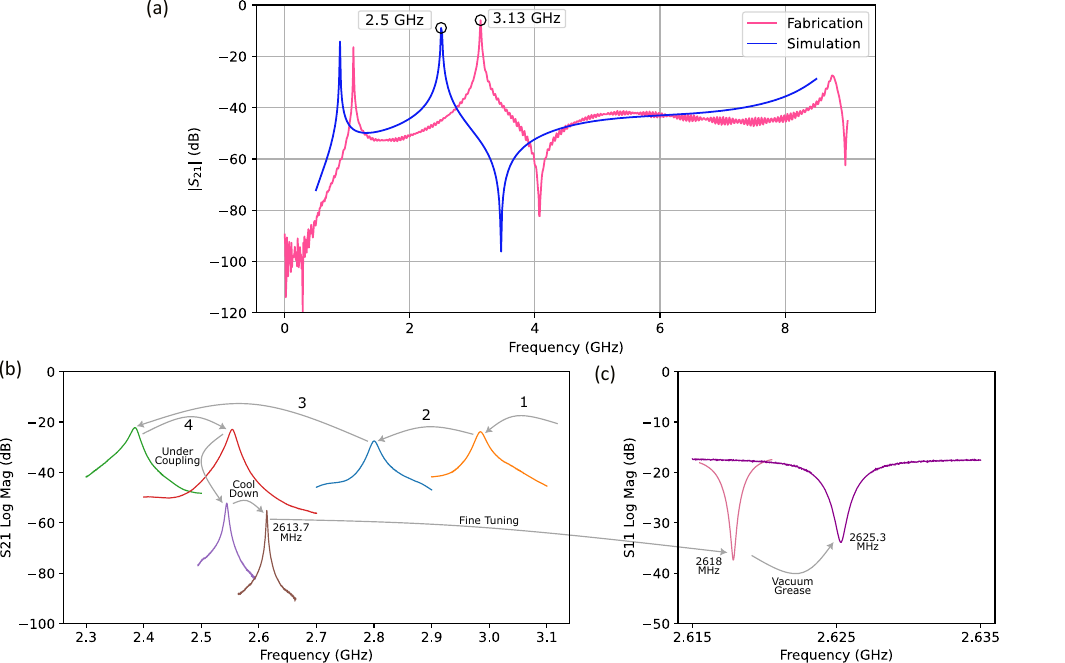}
    \caption{(a) Simulated transmission spectrum of LGR with two ports in COMSOL Multiphysics and measured spectrum after first fabrication. The peaks circled in the spectra correspond to the magnetic field mode with the center loop as the dense region. (b) LGR resonance tuning by removing layers from the outer walls (1-3) to decrease the gap and from the gaps (4) to increase the gap of the LGR. Under-coupling and cool down to 30\,mK improves the quality factor while the latter shifts the resonance frequency as well. (c) Reflection spectrum of the LGR with a single port after fine-tuning the resonance by slowly removing thin layers from the gap.}
    \label{fig:cavityTuning}
\end{figure}


The LGR is simulated in COMSOL Multiphysics with a target to drive the ground state spin transition $\vert2_g\rangle \leftrightarrow \vert4_g\rangle$ at about 2.5\,GHz frequency. It is capacitively coupled to input and output antenna pins. The resonance is observed in the transmission profile as shown in Fig.~\ref{fig:cavityTuning} (a). The second peak (circled) has the relevant field profile needed to interact with the crystal placed in the central loop.

After initial fabrication, the resonant frequency was 3.13\,GHz as shown in Fig.~\ref{fig:cavityTuning} (a). We iteratively tuned the resonance down by removing layers from the outer walls of the LGR and up by removing layers from the gap (see Fig.~\ref{fig:cavityTuning}) as shown in Figs.~\ref{fig:cavityTuning} (b) and (c).  
During several stages of cavity frequency tuning, we realized that the excited state spin transition has better noise properties for transduction, since the microwave field only interacts with the optically excited state ions within the optical mode volume. Therefore, using the excited state spin transition avoids the interaction of microwave field with spectator ions which are not in the optical mode volume but in the microwave mode volume. If ground state spin transition is employed, these spectator ions interact with the microwave field and create noise in the system~\cite{xie-2025, fernandez-gonzalvo-2019}.

Therefore, we tune the LGR resonance at 2613.7\,MHz, close to the excited state spin transition $\vert1_e\rangle \leftrightarrow \vert2_e\rangle$ at frequency 2623.3\,MHz. To improve the extrinsic coupling of the LGR, we increase the length of the pin and place it closer to the gap. Simultaneously, we also decrease the length of the output antenna, which results in under-coupling in the transmission profile. After cooling down the LGR setup, we measure the resonance offset to compensate for, while fine-tuning the resonance. Later, we switch to single-port reflection measurements to get rid of any losses from the output port (Fig.~\ref{fig:cavityTuning} (c)).

After fine-tuning, we arrive at 2618\,MHz. Then we apply vacuum grease to hold and thermalize the crystal, which shifted the resonance frequency to 2625.3\,MHz and broadened the linewidth. This is the final resonance frequency used throughout the experiments. The LGR resonance is repeatedly achieved by applying consistent torque on each of the screws.

\clearpage

\section{Optical Detuning }

\begin{figure}
    \centering
    \includegraphics[width=\linewidth]{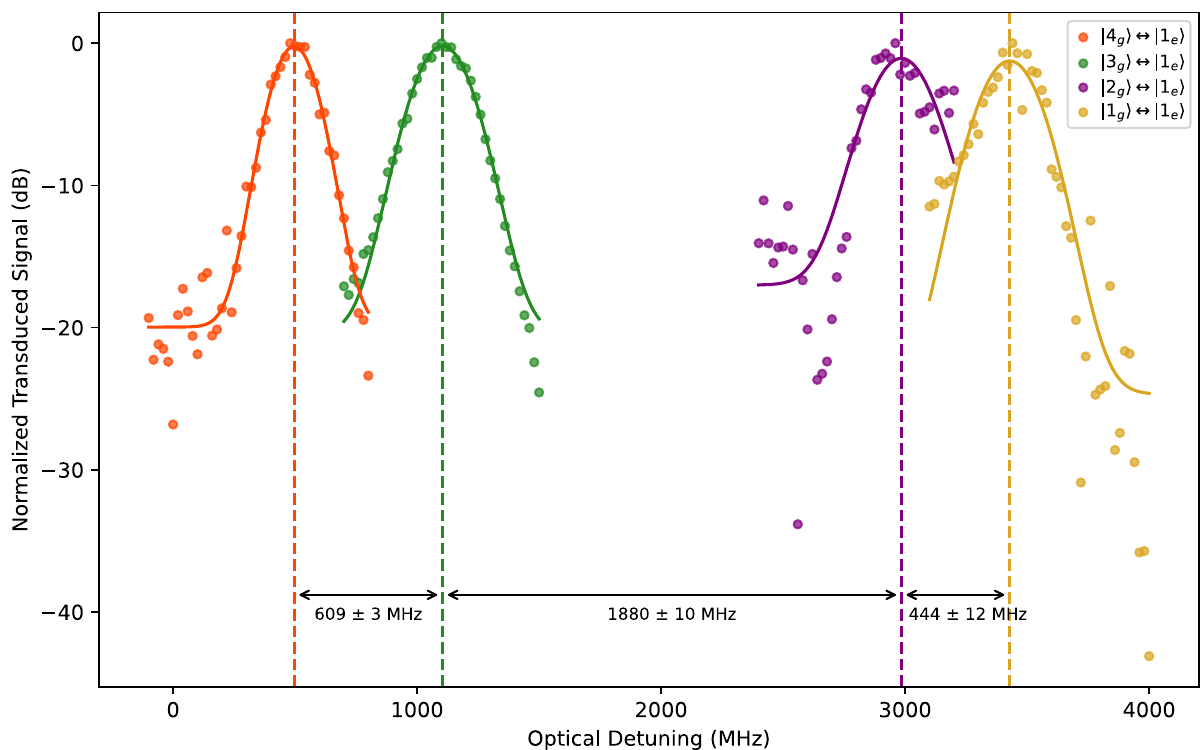}
    \caption{Transduced signals with varying pump detuning around different optical transitions. Each color data points are normalized with respect to the respective peak value. Gaussian fitting of each color set is shown with the solid line. The separation between adjacent fitted peaks are shown and resembles that of the ground state splittings in Yb:YSO. FWHM of the gaussian fits are about 400\,MHz.}
    \label{fig:opticalDetuning}
\end{figure}

Optical detuning of the pump pulse is varied across the ground states to $\vert1_e\rangle$ excited state. It reveals four peaks in the transduction spectrum with varied pump detuning corresponding to each of the transitions as shown in Fig.~\ref{fig:opticalDetuning}. The experiment is performed in four segments, shown with different colours in the figure. For each experiment the laser is locked at the same frequency but shifted near the particular transition using the phase modulator. This provides a single frequency reference for all the measurements. 

Each set of data points are fitted to a gaussian profile shown with the solid line. The separation between the peaks and FWHM of the gaussian curves fairly matches the ground state splittings and optical inhomogeneous broadening (in Fig.~\ref{fig:4g_1e_optical_depth_fit}) when considered with the uncertainty due to the 20\,MHz spacing between each data point. The data is more scattered for low efficiency transduction or for weak local oscillator power (due to high optical depth) used for the heterodyne detection. The signal is normalized to focus on the splittings rather than the peak efficiency, which is already presented in the Fig.~\ref{fig:eff_model_fit}(b).

\end{document}